\documentclass[reqno]{amsart}
\usepackage{amsmath,amsthm,latexsym,amssymb,amscd}
\usepackage{amsfonts}
\usepackage{graphicx,graphics}
\usepackage[french,spanish,english]{babel}

\font\tit=cmbx10 scaled \magstep3 
\font\aut=cmcsc10 scaled \magstep1 \font\afilia=cmr10 scaled
\font\smcc=cmcsc8

\numberwithin{equation}{section}

\begin{document}

\title[Exact solutions for the Generalized Modified Degasperis$-$Procesi equation]{}

\author[Alvaro Salas and Gonzalo Escobar]{}\label{pagini}

\date{}
\maketitle \setcounter{page}{1}

\centerline{\tit New solutions for the  modified generalized}
\centerline{\tit Degasperis$-$Procesi equation}

\vspace{0.8cm} \centerline{\aut Alvaro H. Salas}
\smallskip
\centerline{ Department of Mathematics} \centerline{\afilia
Universidad de Caldas, Manizales, Colombia.}
 \centerline{\afilia Universidad
Nacional de Colombia, Manizales.}

\centerline{ \emph{email} : asalash2002@yahoo.com} \vspace{0.8cm}
\centerline{\aut Jos\'e Gonzalo Escobar Lugo }
\smallskip
\centerline{ Universidad Libre, Bogot\'a, Colombia.}
\centerline{ Universidad Cooperativa de Colombia, Bogot\'a.}
 \centerline{ \emph{email} : jogoel@gmail.com}

\bigskip
{\narrower \noindent {\smcc Abstract.} \small In this paper we show
some new exact solutions for the generalized modified
 Degasperis$-$Procesi equation ( mDP equation )
$ u_t - u_{xxt} + (b+1)u^2u_x = bu_xu_{xx} + uu_{xxx}.$ \par}

\smallskip
{\narrower \noindent {\itp Keywords and phrases.} \small Extended
tanh method, nonlinear ordinary differential equation, nonlinear
partial differential equation, third order evolution equation,
Degasperis$–$Procesi equation,  DP equation, mDP equation,
Camassa-Holm equation, CH equation, mCH equation, traveling wave
solution, partial differential equation ( PDE ), nonlinear evolution
equation.\par}

\smallskip {\narrower \noindent {\itp 2000 Mathematics Subject
 Classification.} \small 35C05. \par}
\normalsize
\section{Introduction}
Degasperis and Procesi \cite{degas} showed, using the method of
asymptotic integrability, that the PDE
\begin{equation}\label{eq00}
   u_t - u_{xxt} + (b+1)uu_x = bu_xu_{xx} + uu_{xxx}
\end{equation}
cannot be completely integrable unless $b = 2$ or $b = 3$. The case
$b = 2$ is the Camassa– Holm (CH) shallow water equation
\cite{cama}, which is well known to be integrable and to possess
multi-soliton (weak) solutions with peaks, so called multi-peakons.
Degasperis, Holm and Hone \cite{degas1,degas2} proved that the case
$b = 3$, which they called the Degasperis– Procesi (DP) equation, is
also integrable and admits multi-peakon solutions. They found the
two-peakon solution explicitly by direct computation.

In this paper we show some exact solutions for the generalized mDP
equation \cite{liu, wazwaz}
\begin{equation}\label{equ00}
u_t - u_{xxt} + (b+1)u^2u_x = bu_xu_{xx} + uu_{xxx}
\end{equation}
for any $b\neq -1$ and $b\neq -2$ by using three distinct methods:
The Cole-Hopf transformation, the rational hyperbolic ansatz and the
improved tanh-coth method.
\section{The Cole-Hopf transformation}
We seek solutions to (\ref{equ00}) in the form of a generalized
Cole-Hopf transformation \cite{cole}\cite{hopf}
\begin{equation}\label{equ01}
u=A\dfrac{\partial^2}{\partial x^2}\log(1+\varphi(x,t))+B,
\end{equation}
for the special choice
\begin{equation}\label{equ02}
\varphi(x,t)=\exp(\mu x+\lambda t+\delta),\quad (A\neq 0,
\quad\lambda\neq0\quad\text{and}\quad\mu\neq 0.)
\end{equation}
From (\ref{equ01}) and (\ref{equ02}) it follows that solutions  have
the form
\begin{equation}\label{equ03}
u=\frac{A \mu ^2}{2 (1+\cosh (\mu\,x+\lambda\,t))}+B.
\end{equation}
Applying this transformation and defining $\zeta=\exp(k x+\omega
t+\delta)$, we obtain a polynomial equation in the variable $\zeta$.
Equating the coefficients of the different powers of $\zeta$ to
zero, we obtain the following algebraic system :
\begin{itemize}
\item $B\mu ^{3}+\lambda \mu ^{2}-\left( bB^{2}+B^{2}\right) \mu -\lambda =0.
$

\item $B \mu  ^3+\lambda \mu ^2+\left(b  B^2+B^2\right) \mu +\lambda  =0.$

\item $(bA+A)\mu ^{5}-(2AB+2AbB+9B)\mu ^{3}-9\lambda \mu ^{2}-\left(
3bB^{2}+3B^{2}\right) \mu -3\lambda =0.$

\item $(b A+A) \mu ^5+(2 A B+2  A b B+9 B) \mu ^3+9 \lambda  \mu ^2+\left(3
b B^2+3  B^2\right) \mu +3 \lambda  =0.$

\item $\left(b A^2+A^2+5 b A+11  A\right) \mu ^5+(2 A B+2 A  b B+10 B) \mu
^3+10 \lambda  \mu ^2+\left(2 b B^2+2  B^2\right) \mu +2 \lambda
=0.$

\item $\left(b A^2+A^2+5 b A+11  A\right) \mu ^5+(2 A B+2 A  b B+10 B) \mu
^3+10 \lambda  \mu ^2+\left(2 b B^2+2  B^2\right) \mu +2 \lambda
=0$.
\end{itemize}
Solving this system gives following solutions :
  \begin{itemize}
\item $A=-\frac{6(b+2)}{b+1},~B=\frac{2\mu ^{2}-1+b\left( \mu ^{2}-1\right) +
\sqrt{1-b(b+2)\left( \mu ^{4}-1\right) }}{2(b+1)},~\lambda
=-\frac{1}{2}\mu \left( b+1-\sqrt{1-b(b+2)\left( \mu ^{4}-1\right)
}\right) :$
\begin{eqnarray*}
u_1(x,t) &=&\frac{2\mu ^{2}-1+b\left( \mu ^{2}-1\right)
+\sqrt{1-b(b+2)\left( \mu
^{4}-1\right) }}{2(b+1)}- \\
&&\frac{6(b+2)\mu ^{2}}{2(b+1)\left( 1+\cosh \left( \mu\,
x-\frac{1}{2}\left( b+1-\sqrt{1-b(b+2)\left( \mu ^{4}-1\right)
}\right) \mu\, t\right) \right) }.
\end{eqnarray*}

\item $A=-\frac{6(b+2)}{b+1},~B=\frac{b\mu ^{2}+2\mu ^{2}-1-b-\sqrt{
1-b(b+2)\left( \mu ^{4}-1\right) }}{2(b+1)},~\lambda
=-\frac{1}{2}\mu \left( b+1+\sqrt{1-b(b+2)\left( \mu ^{4}-1\right)
}\right) :$
\begin{eqnarray*}
&&u_2(x,t)=\frac{b\mu ^{2}+2\mu ^{2}-1-b-\sqrt{1-b(b+2)\left( \mu
^{4}-1\right) }}{
2(b+1)}- \\
&&\frac{6(b+2)\mu ^{2}}{2(b+1)\left( 1+\cosh \left( \mu\,
x-\frac{1}{2}\left( b+1+\sqrt{1-b(b+2)\left( \mu ^{4}-1\right)
}\right) \mu\, t\right) \right) }.
\end{eqnarray*}
\end{itemize}

\section{The rational hyperbolic ansatz}
According this method, we seek solutions in the form
\begin{equation}\label{equ04}
u(x,t)=u(\xi)=\dfrac{a_0+a_1\sinh(\xi)+a_2\cosh(\xi)}{1+c_1\sinh(\xi)+c_2\cosh(\xi)},
\end{equation}
where $\xi=x+\lambda t$ and $\lambda$, $a_0$, $a_1$, $a_2$, $c_1$
and  $c_2$ are constants to be determined. Changing the hyperbolic
functions $\sinh(\xi)$ and $\cosh(\xi)$ to their $\exp$ form and
substituting the resulting expression into (\ref{equ00}), we obtain
a polynomial equation in the variable $\zeta=\exp(\xi)$. Equating
the coefficients of the powers of $\zeta$ to zero, we obtain an
algebraic system in the variables $\lambda$, $a_0$, $a_1$, $a_2$,
$c_1$ and  $c_2$. After solving this system, we get the following
solutions.
\begin{itemize}
\item $\lambda =-\frac{b}{2},\,\,a_{0}=-\frac{3b+5}{b+1},\,\,a_{1}=0,\,
\,a_{2}=\frac{1}{b+1},\,\,c_{1}=0,\,\,c_{2}=1$ :
\begin{equation*}
u_{3}(x,t)=-\frac{3b+5-\cosh \left( x-\frac{bt}{2}\right)
}{(b+1)\left( 1+\cosh \left( x-\frac{bt}{2}\right) \right) }.
\end{equation*}

\item $\lambda =-\frac{b}{2},\,\,a_{0}=-\frac{3b+5}{b+1},\,\,a_{1}=0,\,
\,a_{2}=-\frac{1}{b+1},\,\,c_{1}=0,\,\,c_{2}=-1$ :
\begin{equation*}
u_{4}(x,t)=-\frac{3b+5+\cosh \left( x-\frac{bt}{2}\right)
}{(b+1)\left( 1-\cosh \left( x-\frac{bt}{2}\right) \right) }.
\end{equation*}

\item $\lambda =-\frac{b}{2}-1,\,\,a_{0}=-\frac{3(b+2)}{b+1}
,\,\,a_{1}=0,\,\,a_{2}=0,\,\,c_{1}=0,\,\,c_{2}=-1$ :
\begin{equation*}
u_{5}(x,t)=-\frac{3(b+2)}{(b+1)\left( 1-\cosh \left(
x-(1+\frac{b}{2} )t\right) \right) }.
\end{equation*}

\item $\lambda =-\frac{b}{2}-1,\,\,a_{0}=-\frac{3(b+2)}{b+1}
,\,\,a_{1}=0,\,\,a_{2}=0,\,\,c_{1}=0,\,\,c_{2}=1$ :
\begin{equation*}
u_{6}(x,t)=-\frac{3(b+2)}{(b+1)\left( 1+\cosh \left(
x-(1+\frac{b}{2} )t\right) \right) }.
\end{equation*}

\item $\lambda =-\frac{b}{2},\,\,a_{0}=-\frac{3b+5}{b+1},\,\,a_{1}=-\frac{
\sqrt{(b+1)^{2}a_{2}^{2}-1}}{b+1},\,\,c_{2}=a_{2}(b+1),\,\,c_{1}=-\sqrt{
(b+1)^{2}a_{2}^{2}-1}$ :
\begin{equation*}
u_{7}(x,t)=\frac{-3b-5+(b+1)\cosh \left( x-\frac{bt}{2}\right)
a_{2}-\sqrt{ (b+1)^{2}a_{2}^{2}-1}\sinh \left( x-\frac{bt}{2}\right)
}{(b+1)\left( 1- \sqrt{(b+1)^{2}a_{2}^{2}-1}\sinh \left(
x-\frac{bt}{2}\right) +(b+1)a_{2}\cosh \left( x-\frac{bt}{2}\right)
\right) }.
\end{equation*}

\item $\lambda =-\frac{b}{2},\,\,a_{0}=-\frac{3b+5}{b+1},\,\,a_{1}=\frac{
\sqrt{(b+1)^{2}a_{2}^{2}-1}}{b+1},\,\,\,c_{2}=a_{2}(b+1),\,\,c_{1}=\sqrt{
(b+1)^{2}a_{2}^{2}-1}$ :
\begin{equation*}
u_{8}(x,t)=\frac{-3b-5+(b+1)\cosh \left( x-\frac{bt}{2}\right)
a_{2}+\sqrt{ (b+1)^{2}a_{2}^{2}-1}\sinh \left( x-\frac{bt}{2}\right)
}{(b+1)\left( 1+ \sqrt{(b+1)^{2}a_{2}^{2}-1}\sinh \left(
x-\frac{bt}{2}\right) +(b+1)a_{2}\cosh \left( x-\frac{bt}{2}\right)
\right) }.
\end{equation*}

\item $\lambda =-\frac{b}{2}-1,\,\,a_{0}=-\frac{3(b+2)}{b+1}
,\,\,a_{1}=0,\,\,a_{2}=0,\,\,c_{1}=\sqrt{c_{2}^{2}-1}$ :
\begin{equation*}
u_{9}(x,t)=-\frac{3(b+2)}{(b+1)\left( 1+\sqrt{c_{2}^{2}-1}\sinh
\left( x-(1+ \frac{b}{2})t\right) +c_{2}\cosh \left(
x-(1+\frac{b}{2})t\right) \right) }.
\end{equation*}

\item $\lambda =-\frac{b}{2}-1,\,\,a_{0}=-\frac{3(b+2)}{b+1}
,\,\,a_{1}=0,\,\,a_{2}=0,\,\,c_{1}=-\sqrt{c_{2}^{2}-1}$ :
\begin{equation*}
u_{10}(x,t)=\frac{3(b+2)}{(b+1)\left( 1+\sqrt{c_{2}^{2}-1}\sinh
\left( x-(1+ \frac{b}{2})t\right) -c_{2}\cosh \left(
x-(1+\frac{b}{2})t\right) \right) }.
\end{equation*}
\end{itemize}

\section{The improved tanh-coth method}
The traveling wave transformation
\begin{equation}\label{9}
\xi=x+\lambda t
\end{equation}
reduces (\ref{eq00}) to the nonlinear ode
\begin{equation}\label{equ05}
(b+1) u'(\xi ) u(\xi
   )^2-u'''(\xi ) u(\xi )-\lambda
   u'''(\xi)+\lambda  u'(\xi )-b
   u'(\xi ) u''(\xi )=0.
\end{equation}
We seek solutions to (\ref{equ05}) in the form
\begin{equation}\label{equ06}
u(\xi)=a_0+\sum _{i=1}^m \left(a_i \varphi
   ^i+c_i
   \varphi ^{-i}\right),\\\\
\end{equation}
where $m$ is a positive integer that will be determined and
$\varphi=\varphi(\xi)$ satisfies the Riccati equation
\begin{equation}\label{equ07}
\varphi'(\xi)=\alpha+\beta \varphi(\xi)+\gamma \varphi(\xi)^2,
\end{equation}
where $\alpha$, $\beta$, $\gamma$ are constants. The solutions of
(\ref{equ07}) are given by:
\begin{enumerate}
\item
If $\alpha=0$ and $\beta\neq 0$,
\begin{equation}\label{13}
\varphi(\xi)=\dfrac{\beta}{-\gamma+\beta e^{-\beta \xi}}.
\end{equation}
\item
If $\beta=\alpha=0$ and $\gamma\neq 0$,
\begin{equation}\label{16}
\varphi(\xi)=-\dfrac{1}{\gamma \xi}.
\end{equation}
\item
If $\gamma=0$ and $\beta\neq 0$,
\begin{equation}\label{15}
\varphi(\xi)=\dfrac{-\alpha+\beta e^{\beta \xi}}{\beta}.
\end{equation}
\item
If $\beta=0$ and $\alpha\gamma\neq 0$,
\begin{equation}\label{14}
\begin{cases}
\varphi(\xi)=\dfrac{\sqrt{\alpha \gamma}}{\gamma}\tan(\sqrt{\alpha
\gamma}\xi), & \alpha>0,\gamma>0\\
\varphi(\xi)=\dfrac{\sqrt{\alpha \gamma}}{\gamma}\tanh(\sqrt{\alpha
\gamma}\xi), & \alpha>0,\gamma<0\\
\varphi(\xi)=\dfrac{\sqrt{-\alpha
\gamma}}{\gamma}\tanh(-\sqrt{-\alpha \gamma}\xi), &
\alpha<0,\gamma>0\\
\varphi(\xi)=\dfrac{\sqrt{\alpha \gamma}}{\gamma}\tan(-\sqrt{\alpha
\gamma}\xi), & \alpha<0,\gamma<0.
\end{cases}
\end{equation}
\item If $\beta\neq 0$ and $\beta^2=4\alpha\gamma$,
\begin{equation}\label{18}
\varphi(\xi)=-\frac{2 \alpha  (\beta  \xi
   +2)}{\beta ^2 \xi }
\end{equation}

\item If $\beta^2<4 \alpha\gamma$,
\begin{equation}\label{16a}
\varphi(\xi)=\frac{\sqrt{4 \alpha  \gamma
   -\beta ^2} \tan
   \left(\frac{1}{2} \sqrt{4
   \alpha  \gamma -\beta ^2}
   \xi \right)-\beta }{2
   \gamma }.
\end{equation}

\item If $\beta^2>4 \alpha\gamma$ and $\gamma\neq 0$,
\begin{equation}\label{16b}
\varphi(\xi)=\frac{\sqrt{\beta ^2-4 \alpha
   \gamma } \tanh
   \left(\frac{1}{2}
   \sqrt{\beta ^2-4 \alpha
   \gamma } \xi \right)-\beta
   }{2 \gamma }.
\end{equation}
\end{enumerate}
Substituting (\ref{equ06}), along with (\ref{equ07}) into the left
hand of (\ref{equ05}) and collecting all terms with the same power
in $\varphi(\xi)$,  we get a polynomial in the variable
$\varphi=\varphi(\xi)$. This polynomial has the form
\begin{equation}\label{eq05}
p\varphi^{3m+1}(\xi)+q\varphi^{2m+1}(\xi)+r\varphi^{m+3}(\xi)+\text{lower
degree terms}
\end{equation}
With $m$ determined, we equate each coefficient of the polynomial to
zero. This will give an overdetermined  system of algebraic
equations involving the parameters $\lambda$, $a_0$, $a_i$ and $c_i$
($i = 0,\ldots,m$).  Having determined these parameters,  and using
(\ref{equ06}) and (\ref{13})-(\ref{16b}), we obtain analytic
solutions $u(x, t)$ in a closed form.\\ To determine the parameter
$m$, we usually balance the  terms of highest order in the resulting
polynomial. In our case, we determine $m$ from equations
$3m+1=2m+1$, $3m+1=m+1$, $2m+1=m+3$. This gives $m=0$ or $m=2$. We
define $m=2$ to avoid trivial solutions, so solutions have the form
\begin{equation}\label{eq06}
u(\xi)=a_0+a_1\varphi(\xi)+\dfrac{c_1}{\varphi(\xi)}+a_2\varphi^2(\xi)+\dfrac{c_2}{\varphi^2(\xi)}.
\end{equation}
After solving the algebraic system , following solutions are
obtained : \vspace{0.5cm}\newline \textbf{First Case} : $\beta \neq
0$ and $\beta ^{2}=4\alpha \gamma .$
\newline
\newline
$\bullet $ $a_{0}=\frac{3(b+2)\beta ^{2}}{2(b+1)}-1$, $\,\,a_{1}=0$,
$ \,\,a_{2}=0$, $\,\,c_{1}=\frac{6(b+2)\alpha \beta }{b+1}$,
$\,\,c_{2}=\frac{ 6(b+2)\alpha ^{2}}{b+1}$, $\,\,\lambda =-b-1:$
\begin{equation*}
u_{11}(x,t)=\frac{6(b+2)\beta ^{2}}{(b+1)(\beta \xi +2)^{2}}-1.
\end{equation*}
\textbf{Second Case} : $\alpha =0$ and $\beta \neq 0.$\newline
\newline
$\bullet $ $\,\,a_{0}=-\frac{-b\beta ^{2}-2\beta
^{2}+b+\sqrt{1-b(b+2)\left( \beta ^{4}-1\right) }+1}{2(b+1)}$,
$\,a_{1}=\frac{6(b+2)\beta \gamma }{b+1}$,
$\,\,a_{2}=\frac{6(b+2)\gamma ^{2}}{b+1}$,
$\,\,c_{1}=0$,$\,\,c_{2}=0$,
\newline
$\lambda =\frac{1}{2}\left( -b-\sqrt{1-b(b+2)\left( \beta
^{4}-1\right) } -1\right) :$
\begin{eqnarray*}
\,u_{12}(x,\,\,t) &=&\frac{1}{2(b+1)}\left( (b+2)\beta
^{2}-b-1-\sqrt{
1-b(b+2)\left( \beta ^{4}-1\right) }\right) + \\
&&\frac{6(b+2)\gamma \beta ^{2}}{b+1}\left( \frac{1}{e^{-\beta \xi
}\beta -\gamma }+\frac{\gamma }{\left( e^{-\beta \xi }\beta -\gamma
\right) ^{2}} \right) .
\end{eqnarray*}
\newline
$\bullet $ $\,a_{0}=\frac{2\beta ^{2}+b\left( \beta ^{2}-1\right)
+\sqrt{ 1-b(b+2)\left( \beta ^{4}-1\right)
}-1}{2(b+1)}$,$\,a_{1}=\frac{6(b+2)\beta \gamma
}{b+1}$,$\,\,a_{2}=\frac{6(b+2)\gamma ^{2}}{b+1}$, $\,c_{1}=0$,$
\,\,c_{2}=0$,$\,$\newline $\,\ \lambda =\frac{1}{2}\left(
-b+\sqrt{1-b(b+2)\left( \beta ^{4}-1\right) } -1\right) :$
\begin{eqnarray*}
\,u_{13}(x,\,\,t) &=&\frac{1}{2(b+1)}\left( (b+2)\beta
^{2}-b-1+\sqrt{
1-b(b+2)\left( \beta ^{4}-1\right) }\right) + \\
&&\frac{6(b+2)\gamma \beta ^{2}}{b+1}\left( \frac{1}{e^{-\beta \xi
}\beta -\gamma }+\frac{\gamma }{\left( e^{-\beta \xi }\beta -\gamma
\right) ^{2}} \right) .
\end{eqnarray*}
\textbf{Third Case} : $\beta =0$ and $\alpha \gamma \neq 0:$
\newline
\newline
$\bullet $ $a_{0}=-\frac{-8\alpha \gamma b+b-16\alpha \gamma +\sqrt{
b(b+2)\left( 1-256\alpha ^{2}\gamma ^{2}\right)
+1}+1}{2(b+1)}$,$\,\,a_{1}=0$ ,$\,\,a_{2}=\frac{6(b+2)\gamma
^{2}}{b+1}$,$\,\,c_{1}=0$,$\,\,c_{2}=\frac{ 6(b+2)\alpha
^{2}}{b+1}$,\newline $\lambda =\frac{1}{2}\left(
-b-\sqrt{b(b+2)\left( 1-256\alpha ^{2}\gamma ^{2}\right)
+1}-1\right) :$
\begin{equation*}
u_{14}(x,t)=\frac{-(b+1)-16\alpha \gamma (b+2)-\sqrt{b(b+2)\left(
1-256\alpha ^{2}\gamma ^{2}\right) +1}+48\alpha \gamma (b+2)\csc
^{2}\left( 2 \sqrt{\alpha \gamma }\xi \right) }{2(b+1)}.
\end{equation*}
$\bullet $ $a_{0}=\frac{8\alpha \gamma b-b+16\alpha \gamma +\sqrt{
b(b+2)\left( 1-256\alpha ^{2}\gamma ^{2}\right)
+1}-1}{2(b+1)}$,$\,\,a_{1}=0$ ,$\,\,a_{2}=\frac{6(b+2)\gamma
^{2}}{b+1}$,$\,\,c_{1}=0$,$\,\,c_{2}=\frac{ 6(b+2)\alpha
^{2}}{b+1}$,\newline $\lambda =\frac{1}{2}\left(
-b+\sqrt{b(b+2)\left( 1-256\alpha ^{2}\gamma ^{2}\right)
+1}-1\right) :$
\begin{equation*}
u_{15}(x,t)=\frac{-(b+1)-16\alpha \gamma (b+2)+\sqrt{b(b+2)\left(
1-256\alpha ^{2}\gamma ^{2}\right) +1}+48\alpha \gamma (b+2)\csc
^{2}\left( 2 \sqrt{\alpha \gamma }\xi \right) }{2(b+1)}.
\end{equation*}
$\bullet $ $a_{0}=-\frac{-8\alpha \gamma b+b-16\alpha \gamma +\sqrt{
b(b+2)\left( 1-16\alpha ^{2}\gamma ^{2}\right)
+1}+1}{2(b+1)}$,$\,\,a_{1}=0$,
$\,\,a_{2}=0$,$\,\,c_{1}=0$,$\,\,c_{2}=\frac{6(b+2)\alpha
^{2}}{b+1}$,
\newline
$\lambda =\frac{1}{2}\left( -b-\sqrt{b(b+2)\left( 1-16\alpha
^{2}\gamma ^{2}\right) +1}-1\right) :$
\begin{equation*}
u_{16}(x,t)=\frac{-(b+1)+8\alpha \gamma (b+2)-\sqrt{b(b+2)\left(
1-16\alpha ^{2}\gamma ^{2}\right) +1}+12\alpha \gamma (b+2)\cot
^{2}\left( \sqrt{\alpha \gamma }\xi \right) }{2(b+1)}.
\end{equation*}
$\bullet $ $a_{0}=-\frac{-8\alpha \gamma b+b-16\alpha \gamma +\sqrt{
b(b+2)\left( 1-16\alpha ^{2}\gamma ^{2}\right)
+1}+1}{2(b+1)}$,$\,\,a_{1}=0$, $\,\,a_{2}=\frac{6(b+2)\gamma
^{2}}{b+1}$,$\,\,c_{1}=0$,$\,\,c_{2}=0$,
\newline
$\lambda =\frac{1}{2}\left( -b-\sqrt{b(b+2)\left( 1-16\alpha
^{2}\gamma ^{2}\right) +1}-1\right) :$
\begin{equation*}
u_{17}(x,t)=\frac{-(b+1)+8\alpha \gamma (b+2)-\sqrt{b(b+2)\left(
1-16\alpha ^{2}\gamma ^{2}\right) +1}+12\alpha \gamma (b+2)\tan
^{2}\left( \sqrt{\alpha \gamma }\xi \right) }{2(b+1)}.
\end{equation*}
$\bullet $ $a_{0}=\frac{8\alpha \gamma b-b+16\alpha \gamma +\sqrt{
b(b+2)\left( 1-16\alpha ^{2}\gamma ^{2}\right)
+1}-1}{2(b+1)}$,$\,\,a_{1}=0$,
$\,\,a_{2}=0$,$\,\,c_{1}=0$,$\,\,c_{2}=\frac{6(b+2)\alpha
^{2}}{b+1}$,
\newline
$\lambda =\frac{1}{2}\left( -b+\sqrt{b(b+2)\left( 1-16\alpha
^{2}\gamma ^{2}\right) +1}-1\right) :$
\begin{equation*}
u_{18}(x,t)=\frac{-(b+1)+8\alpha \gamma (b+2)+\sqrt{b(b+2)\left(
1-16\alpha ^{2}\gamma ^{2}\right) +1}+12\alpha \gamma (b+2)\cot
^{2}\left( \sqrt{\alpha \gamma }\xi \right) }{2(b+1)}.
\end{equation*}
$\bullet $ $a_{0}=\frac{8\alpha \gamma b-b+16\alpha \gamma +\sqrt{
b(b+2)\left( 1-16\alpha ^{2}\gamma ^{2}\right)
+1}-1}{2(b+1)}$,$\,\,a_{1}=0$, $\,\,a_{2}=\frac{6(b+2)\gamma
^{2}}{b+1}$,$\,\,c_{1}=0$,$\,\,c_{2}=0$,
\newline
$\lambda =\frac{1}{2}\left( -b+\sqrt{b(b+2)\left( 1-16\alpha
^{2}\gamma ^{2}\right) +1}-1\right) :$
\begin{equation*}
u_{19}(x,t)=\frac{-(b+1)+8\alpha \gamma (b+2)+\sqrt{b(b+2)\left(
1-16\alpha ^{2}\gamma ^{2}\right) +1}+12\alpha \gamma (b+2)\tan
^{2}\left( \sqrt{\alpha \gamma }\xi \right) }{2(b+1)}.
\end{equation*}
\textbf{Fourth Case} : $\Delta \neq 0$, where $\Delta =\beta
^{2}-4\alpha \gamma :$\newline
\newline
$\bullet $ $a_{0}=\frac{24\alpha \gamma +2\Delta +b(12\alpha \gamma
+\Delta -1)-\sqrt{1-b(b+2)\left( \Delta ^{2}-1\right)
}-1}{2(b+1)}$,$\,\,a_{1}=\frac{ 6(b+2)\beta \gamma
}{b+1}$,$\,\,a_{2}=\frac{6(b+2)\gamma ^{2}}{b+1}$,$
\,\,c_{1}=0$,$\,\,c_{2}=0$,\newline $\,\,\lambda =\frac{1}{2}\left(
-b-\sqrt{1-b(b+2)\left( \Delta ^{2}-1\right) }-1\right) :$
\begin{equation*}
u_{20}(x,t)=-\frac{2\Delta (b+2)+b+1+\sqrt{1-b(b+2)\left( \Delta
^{2}-1\right) }}{2(b+1)}+\frac{3(b+2)\Delta }{2(b+1)}\tanh
^{2}\left( \frac{1 }{2}\sqrt{\Delta }\xi \right) .
\end{equation*}
$\bullet $ $a_{0}=\frac{24\alpha \gamma +2\Delta +b(12\alpha \gamma
+\Delta -1)+\sqrt{1-b(b+2)\left( \Delta ^{2}-1\right)
}-1}{2(b+1)}$,$\,\,a_{1}=\frac{ 6(b+2)\beta \gamma
}{b+1}$,$\,\,a_{2}=\frac{6(b+2)\gamma ^{2}}{b+1}$,$
\,\,c_{1}=0$,$\,\,c_{2}=0$,\newline $\,\,\lambda =\frac{1}{2}\left(
-b+\sqrt{1-b(b+2)\left( \Delta ^{2}-1\right) }-1\right) :$
\begin{equation*}
u_{21}(x,t)=-\frac{2\Delta (b+2)+b+1-\sqrt{1-b(b+2)\left( \Delta
^{2}-1\right) }}{2(b+1)}+\frac{3(b+2)\Delta }{2(b+1)}\tanh
^{2}\left( \frac{1 }{2}\sqrt{\Delta }\xi \right) .
\end{equation*}
$\bullet $ $a_{0}=\frac{24\alpha \gamma +2\Delta +b(12\alpha \gamma
+\Delta -1)+\sqrt{1-b(b+2)\left( \Delta ^{2}-1\right)
}-1}{2(b+1)}$,$\,\,a_{1}=0$,$
\,\,a_{2}=0$,$\,\,c_{1}=\frac{6(b+2)\alpha \beta
}{b+1}$,$\,\,c_{2}=\frac{ 6(b+2)\alpha ^{2}}{b+1}$,\newline
$\,\,\lambda =\frac{1}{2}\left( -b+\sqrt{1-b(b+2)\left( \Delta
^{2}-1\right) }-1\right) :$
\begin{eqnarray*}
u_{22}(x,t) &=&\frac{1}{2(b+1)}((\Delta +12\alpha \gamma
)(b+2)-(b+1)+\sqrt{
1-b(b+2)\left( \Delta ^{2}-1\right) })- \\
&&\frac{12(b+2)\alpha \gamma \left( \beta ^{2}+\sqrt{\Delta }\tanh
\left( \frac{1}{2}\sqrt{\Delta }\xi \right) \beta -2\alpha \gamma
\right) }{ (b+1)\left( \beta +\sqrt{\Delta }\tanh \left(
\frac{1}{2}\sqrt{\Delta }\xi \right) \right) ^{2}}.
\end{eqnarray*}
$\bullet $ $a_{0}=\frac{24\alpha \gamma +2\Delta +b(12\alpha \gamma
+\Delta -1)-\sqrt{1-b(b+2)\left( \Delta ^{2}-1\right)
}-1}{2(b+1)}$,$\,\,a_{1}=0$,$
\,\,a_{2}=0$,$\,\,c_{1}=\frac{6(b+2)\alpha \beta
}{b+1}$,$\,\,c_{2}=\frac{ 6(b+2)\alpha ^{2}}{b+1}$,\newline
$\,\,\lambda =\frac{1}{2}\left( -b-\sqrt{1-b(b+2)\left( \Delta
^{2}-1\right) }-1\right) :$
\begin{eqnarray*}
u_{23}(x,t) &=&\frac{1}{2(b+1)}\left( (\Delta +12\alpha \gamma
)(b+2)-(b+1)-
\sqrt{1-b(b+2)(\Delta ^{2}-1)}\right) - \\
&&\frac{12(b+2)\alpha \gamma \left( \beta ^{2}+\sqrt{\Delta }\tanh
\left( \frac{1}{2}\sqrt{\Delta }\xi \right) \beta -2\alpha \gamma
\right) }{ (b+1)\left( \beta +\sqrt{\Delta }\tanh \left(
\frac{1}{2}\sqrt{\Delta }\xi \right) \right) ^{2}}.
\end{eqnarray*}

\section{Conclusions}
In this paper, by using three distinct methods and the help of a
symbolic computation engine, we obtain some exact solutions for the
equation (\ref{eq00}). The methods we employ certainly work well for
a large class of very interesting nonlinear equations. The main
advantage of these methods is that their capability of greatly
reducing the size of computational work compared to existing
techniques such as the pseudo spectral method, the inverse
scattering method, Hirota's bilinear method, and the truncated
Painlev\'e expansion. We think that some of the solutions we
obtained are new in the literature.

\end{document}